\documentclass[a4paper]{jpconf}
\usepackage{graphicx}
\begin{document}
\title{The Phase Diagram of Two Color QCD}

\author{Simon Hands$^1$, Seamus Cotter$^2$, Pietro Giudice$^1$ and \\Jon-Ivar
Skullerud$^2$}

\address{$^1$ Physics Department, College of Science, Swansea University,
Singleton Park, \\Swansea SA2 8PP, UK}

\address{$^2$ Department of Mathematical Physics, National University of Ireland
Maynooth, \\Maynooth, County Kildare, Ireland}

\ead{s.hands@swan.ac.uk}

\begin{abstract}
I present recent results from lattice simulations of SU(2) gauge theory with
$N_f=2$ Wilson quark flavors, at non-zero quark chemical potential $\mu$. The
thermodynamic equation of state is discussed along with the nature of the high
density matter which forms. It is conjectured that deconfinement may mean
different things for bulk and Fermi surface phenomena.
\end{abstract}

\section{Why Two Colors?}
In this talk I will give an update on our project to simulate SU(2) gauge theory
with $N_f=2$ quark flavors on cold lattices with quark chemical
potential $\mu\not=0$~\cite{QC2D1,QC2D2,QC2D3}. Because of the Sign Problem,
QC$_2$D offers currently the best prospect of 
using lattice simulations to study gauge theories at non-zero
baryon charge density. In particular, we can explore
the systematics of the lattice approach in this unexplored physical regime.
There is good news: unlike the case of hot QCD, it is possible to perform a
scan in $\mu$ at fixed cutoff, and the primary thermodynamic observable, the
quark number density $n_q\equiv-\partial f/\partial\mu$, as a component of a
conserved current has no quantum corrections. The bad news is
that both UV and IR artifacts are significant and complicated; only
recently have we begun to make progress in disentangling them~\cite{QC2D3}.

A second, more theoretical motivation is that QC$_2$D with $\mu\not=0$ offers
the chance to explore color deconfinement in a new physical r\'egime,
complementary to the more usually studied transition at $T>0$. There
are interesting differences which will challenge us to refine the language
used to discuss this essential feature of non-abelian gauge theories.


\section{The Simulation}

We define our lattice theory with the Wilson plaquette  action 
and unimproved Wilson fermions with hopping
parameter $\kappa$. Chemical potential is introduced via the standard
prescription of weighting forward/backward temporal hops by $e^{\pm\mu a}$.
Reality of the fermion measure follows from ${\rm det}M={\rm
det}\tau_2M^*\tau_2$ where the Pauli matrix $\tau_2$ acts on color.
Our only innovation~\cite{QC2D1} is the introduction of a scalar, isoscalar and
gauge-invariant diquark source term
$jqq\equiv j\kappa(\psi_2^{tr}C\gamma_5\tau_2\psi_1-\bar\psi_1C\gamma_5\tau_2\bar\psi_2^{tr})$, 
where subscripts denote flavor. Setting $j\not=0$ mitigates the IR
fluctuations due to Goldstone modes associated with superfluidity at non-zero
quark density. 
Indeed, $j$ is nothing but a Majorana mass for the quarks.
\begin{figure}[h]
\begin{minipage}{18pc}
\includegraphics[width=18pc]{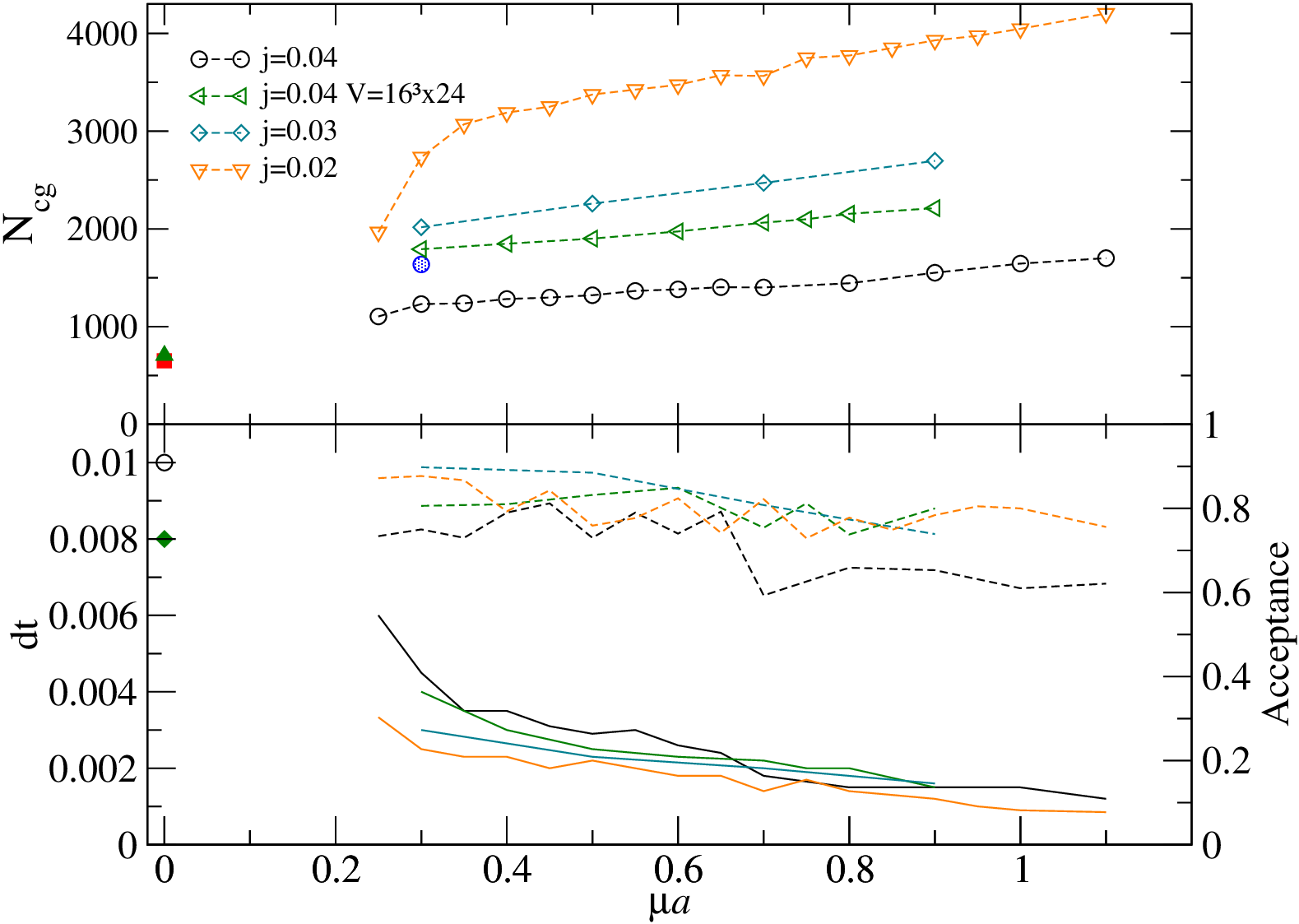}
\caption{\label{fig:effort} Computational effort needed for QC$_2$D.}
\end{minipage}\hspace{2pc}%
\begin{minipage}{18pc}
\includegraphics[width=18pc]{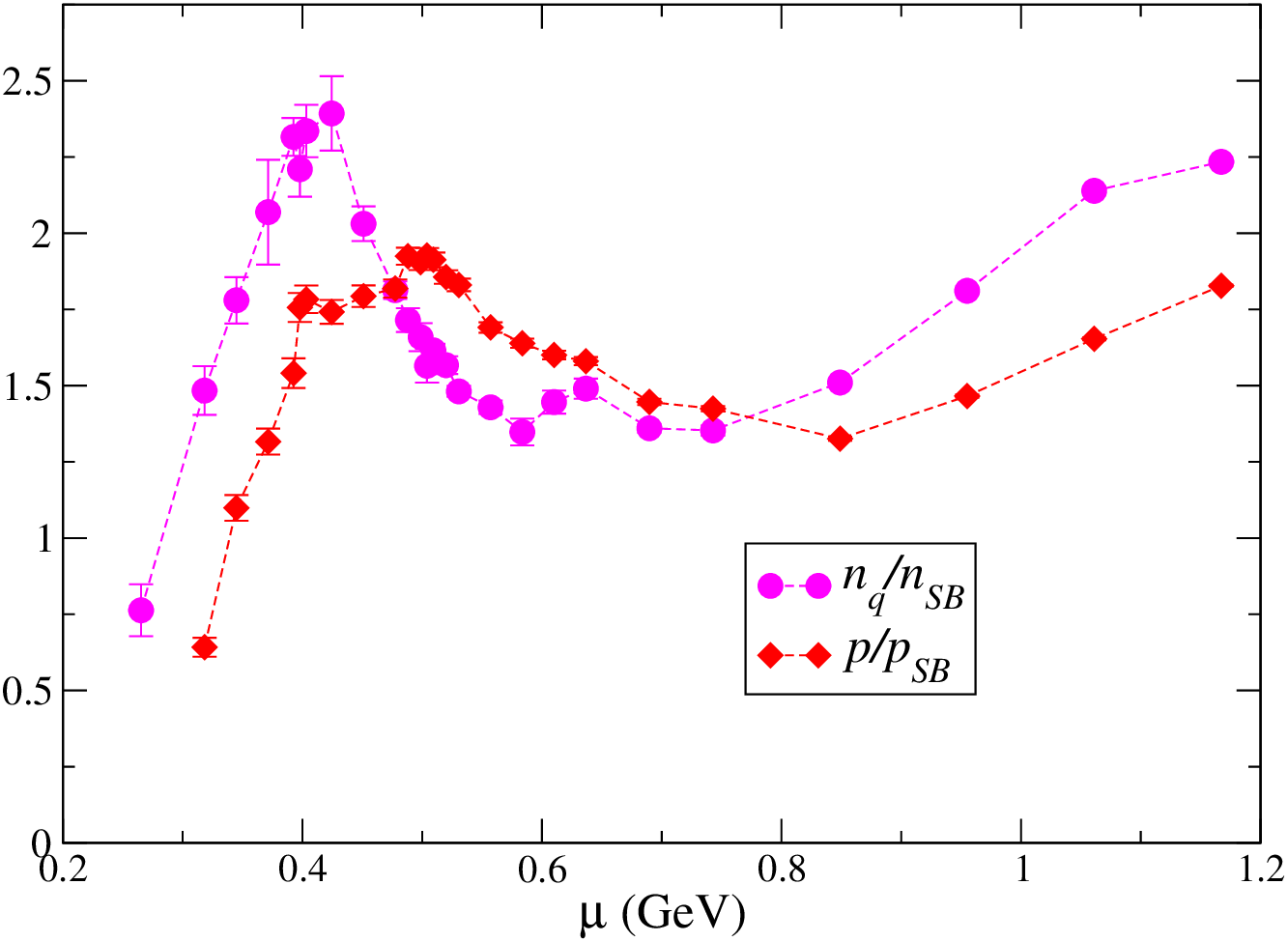}
\caption{\label{fig:nice} $n_q/n_q^{SB}$ and $p/p^{SB}$ for $ja=0.04$ on
$12^3\times24$.}
\end{minipage} 
\end{figure}
Fig.~\ref{fig:effort} shows how the computational effort, measured by both the
number of {\tt congrad} iterations and the HMC timestep $dt$ required for decent
acceptance varies with $\mu$ and $j$. Despite the absence of a Sign Problem, 
studies of dense matter still require an order of magnitude more
cpu than that needed for vacuum QCD.

By now we have accumulated matched ensembles on a coarse $8^3\times16$ lattice with
$a=0.229(3)$fm~\cite{QC2D1} and a finer $12^3\times24$ lattice with
$a=0.178(6)$fm~\cite{QC2D2}, with scale set by assuming the two color
string tension is (440MeV)$^2$. Both lattices have temperature
$T\approx50$MeV, making them ``cold'' in QCD terms. The quark mass is
relatively heavy, corresponding to $m_\pi/m_\rho\approx0.8$ --- 
up to now we have
not considered the chiral limit to be so important, 
because effects due to $m\not=0$ could naively be expected to lie deep
at the bottom of the Fermi Sea. In our most recent simulations~\cite{QC2D3} we
have explored the systematics both of varying the diquark source
$ja=0.04,\ldots,0.02$ and varying $T$, with $N_\tau=24,\ldots,8$ corresponding to 
$T=47,\ldots,141$MeV.

\section{Equation of State}
\label{sec:eos}

\begin{figure}[h]
\begin{minipage}{18pc}
\includegraphics[width=18pc]{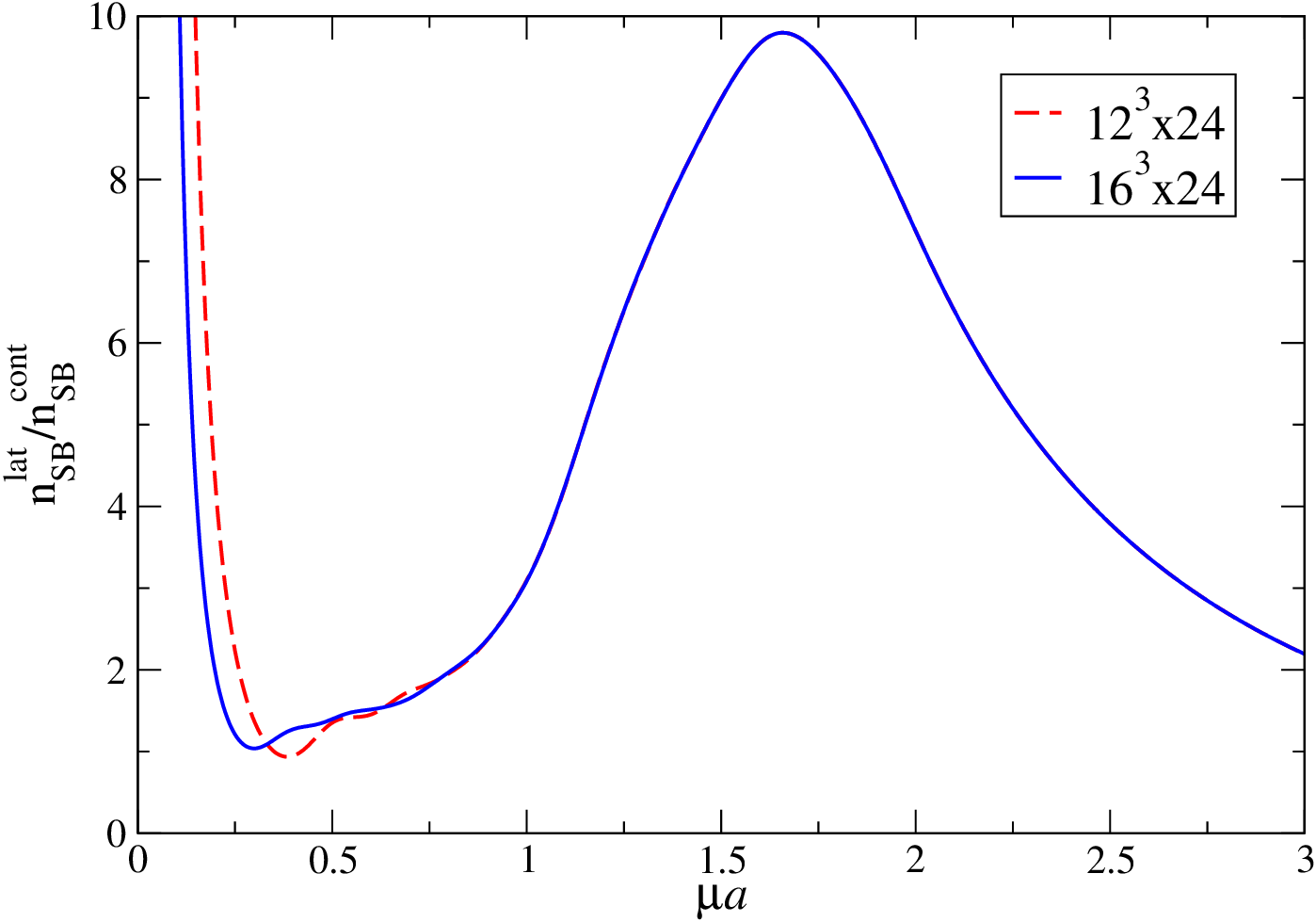}
\caption{\label{fig:freefermion} $3\pi^2n_q^{SB}/N_fN_c\mu^3$ vs. $\mu a$ on two
different lattice volumes.}
\end{minipage}\hspace{2pc}%
\begin{minipage}{18pc}
\includegraphics[width=16pc]{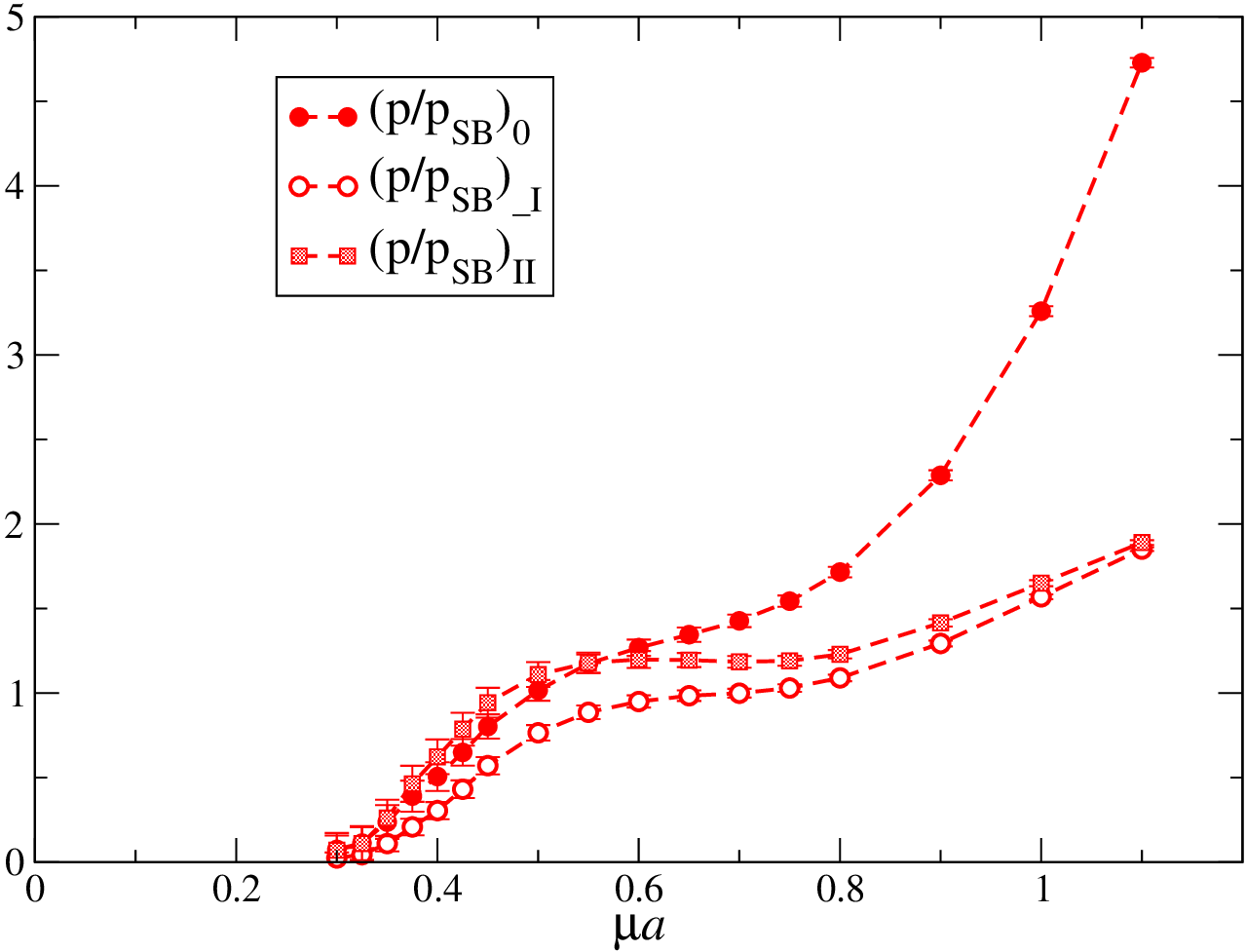}
\caption{\label{fig:jto0} $p/p^{SB}$ on $12^3\times24$ for $j\to0$ using three different
artifact correction schemes.}
\end{minipage} 
\end{figure}

Fig.~\ref{fig:nice} plots $n_q/n_q^{SB}$ as a function of $\mu a$,
calculated on the fine lattice
with $ja=0.04$~\cite{QC2D2}. Here $n_q^{SB}$ denotes the ``Stefan-Boltzmann''
(SB)
result evaluated for free Wilson fermions on the same volume;
taking the ratio is a first step towards correcting for discretisation and
finite volume artifacts. These are considerable, as shown by the
departures from unity in
Fig.~\ref{fig:freefermion}. For $\mu a$ greater than unity, the dominant error
is due to UV artifacts, but for smaller $\mu$ the disparity between the two
curves points to the IR; indeed, the oscillatory behaviour is due to departures
of the Fermi surface from sphericity as $T\to0$ due to the discretisation of
momentum space~\cite{HW}.
\begin{figure}[h]
\begin{minipage}{18pc}
\includegraphics[width=18pc]{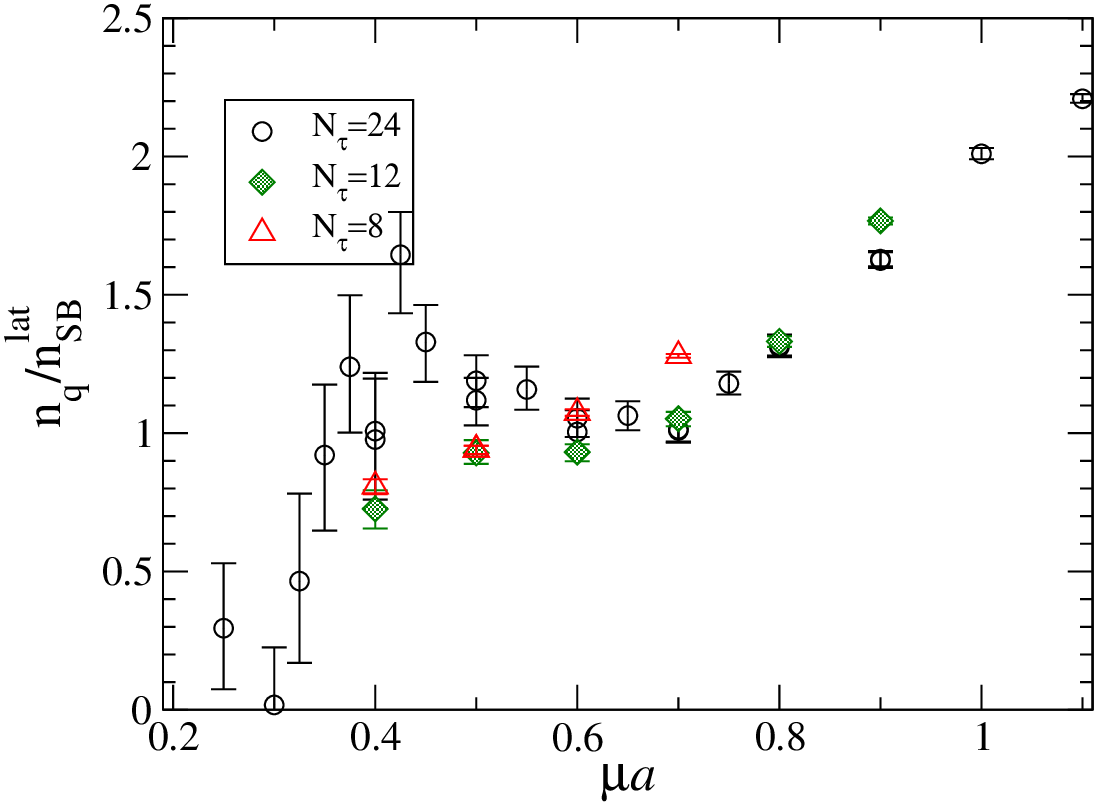}
\caption{\label{fig:nq_j0_T_a} $n_q/n_q^{SB}$ for $j\to0$ using lattice free
fermions.}
\end{minipage}\hspace{2pc}%
\begin{minipage}{18pc}
\includegraphics[width=17pc]{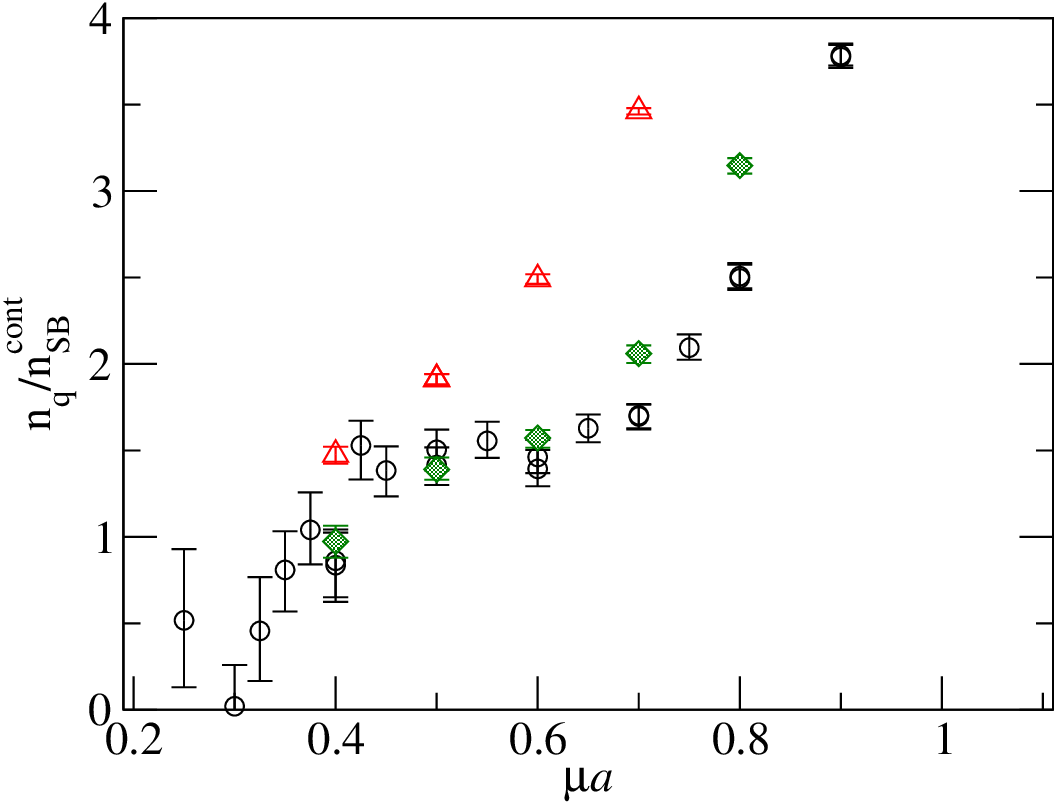}
\caption{\label{fig:nq_j0_T_b} $n_q/n_q^{SB}$ for $j\to0$ using continuum free
fermions.}
\end{minipage} 
\end{figure}
Fig.~\ref{fig:nq_j0_T_a} illustrates the difficulty of disentangling these two
effects, updating Fig.~\ref{fig:nice} for a range of temperatures
with the limit $j\to0$ taken. Fig.~\ref{fig:nq_j0_T_b} presents the same data 
normalised by the continuum SB result $N_fN_c(\mu^3/\pi^2+\mu
T^2)/3$. The apparent peak at $\mu a\approx0.4$, previously
identified as a signal of Bose-Einstein condensation (BEC) of tightly bound 
diquarks~\cite{QC2D1,QC2D2}, is most likely an IR artifact due to the
dip in the dashed red free fermion curve of Fig.~\ref{fig:freefermion}. On the other hand the
difference in vertical scales shows that UV artifacts
still need to be corrected.

Another previously neglected factor is the $j\to0$ limit,
which appears to be a much more important effect for interacting quarks; 
setting $j\not=0$ enhances their tendency to form pairs, in effect promoting BEC
formation which could deform the results. 
Taking all factors into account, we conclude the approximate plateau in the 
range $0.4<\mu a<0.7$ is consistent with $n_q/n_q^{SB}\approx1$, rather than
1.4 as suggested by Fig.~\ref{fig:nice}. The same gross features are also
apparent in a calculation of the pressure $p=\int^\mu n_qd\mu$;
Fig.~\ref{fig:jto0} shows the result of three different approaches to correcting
for lattice artifacts at $T=47$MeV~\cite{QC2D3}.

Figs.~\ref{fig:nq_j0_T_a},\ref{fig:nq_j0_T_b} and \ref{fig:jto0} 
support an interpretation in terms of four distinct regimes
separated by three thresholds. The first is the onset threshold, taking place at 
$\mu_o=m_\pi/2\approx0.32a^{-1}\approx360$MeV, separating the vacuum state from
a non-zero quark density $n_q>0$. In mean field theory this is predicted to be a
second order phase transition~\cite{KSTVZ}, unlike the first order transition
expected in QCD. The non-monotonic behaviour of
$n_q/n_q^{SB}$ above onset due to a BEC, predicted in mean field-theory, and reported in \cite{QC2D1,QC2D2}
is, however, no longer a robust feature of the numerical results; it may well be that
simulations far closer to the chiral limit are needed to expose this behaviour.
Next, at $\mu_Q\approx0.5a^{-1}\approx530$MeV is a crossover to a r\'egime
where both $n_q$ and $p$ are approximately equal to their SB
values, suggestive that a degenerate system of quarks with a well-defined Fermi
momentum $k_F\propto n_q^{1/3}$ and Fermi energy $E_F\approx\mu$ 
has formed. For reasons that will be elaborated we refer to this as the {\it
quarkyonic\/} r\'egime. Finally at $\mu_d\approx0.8a^{-1}\approx850$MeV both
$n_q$ and $p$ start to climb above the SB values. If for the moment we assume
the degenerate quark description still holds, then for $\mu_Q<\mu<\mu_d$ we
have $E_F\approx k_F$ as expected for weakly interacting massless quarks, but
$E_F<k_F$ for $\mu>\mu_d$, implying that the quark matter at very large
densities has a large negative correction to the kinetic energy, ie. it is strongly self-bound.

\section{Order Parameters and Phase Diagram}

\begin{figure}[h]
\begin{minipage}{18pc}
\includegraphics[width=18pc]{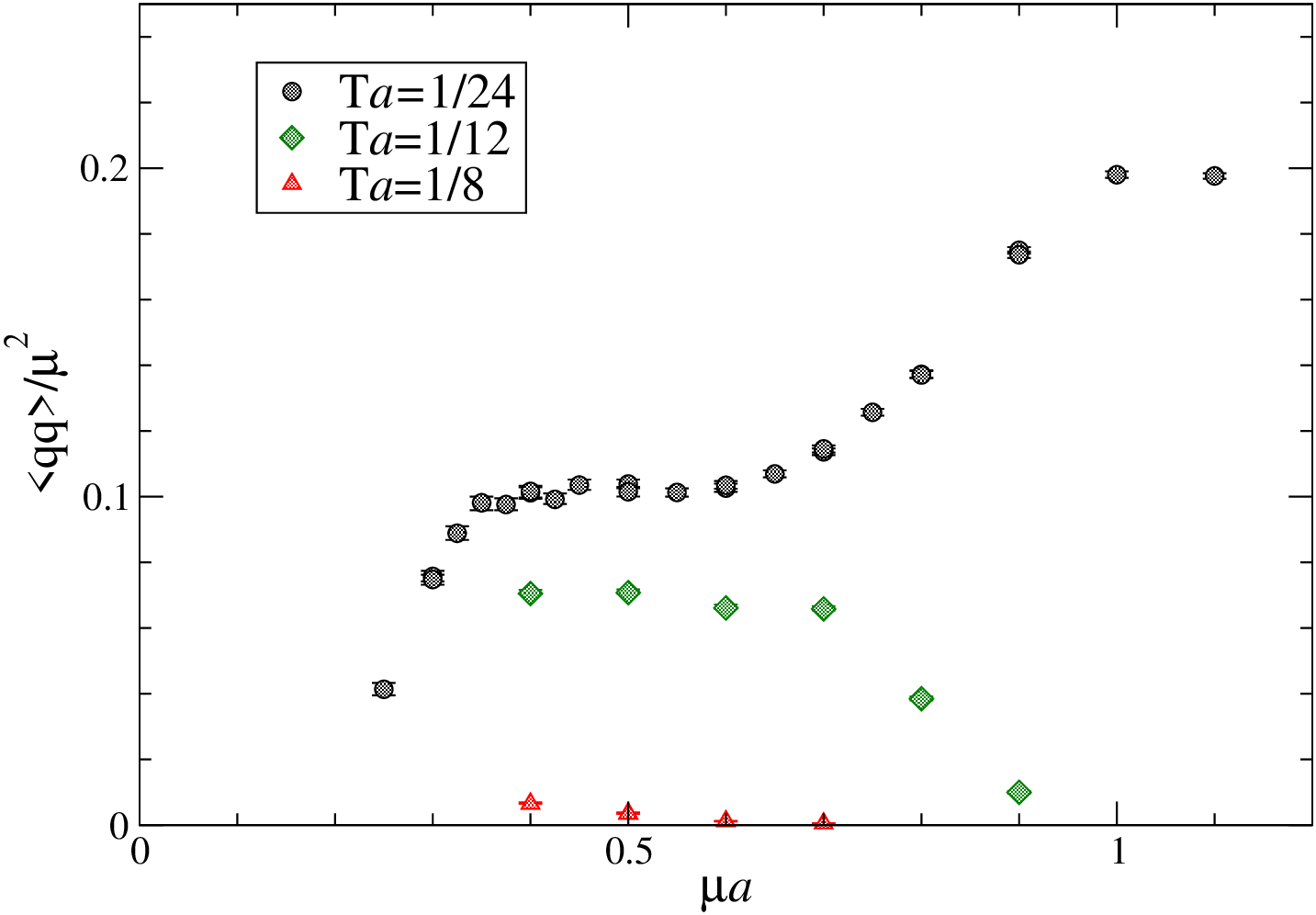}
\caption{\label{fig:qq_j0} $\langle qq\rangle/\mu^2$ for $j\to0$ for various
$T$.}
\end{minipage} 
\begin{minipage}{18pc}
\includegraphics[width=17pc]{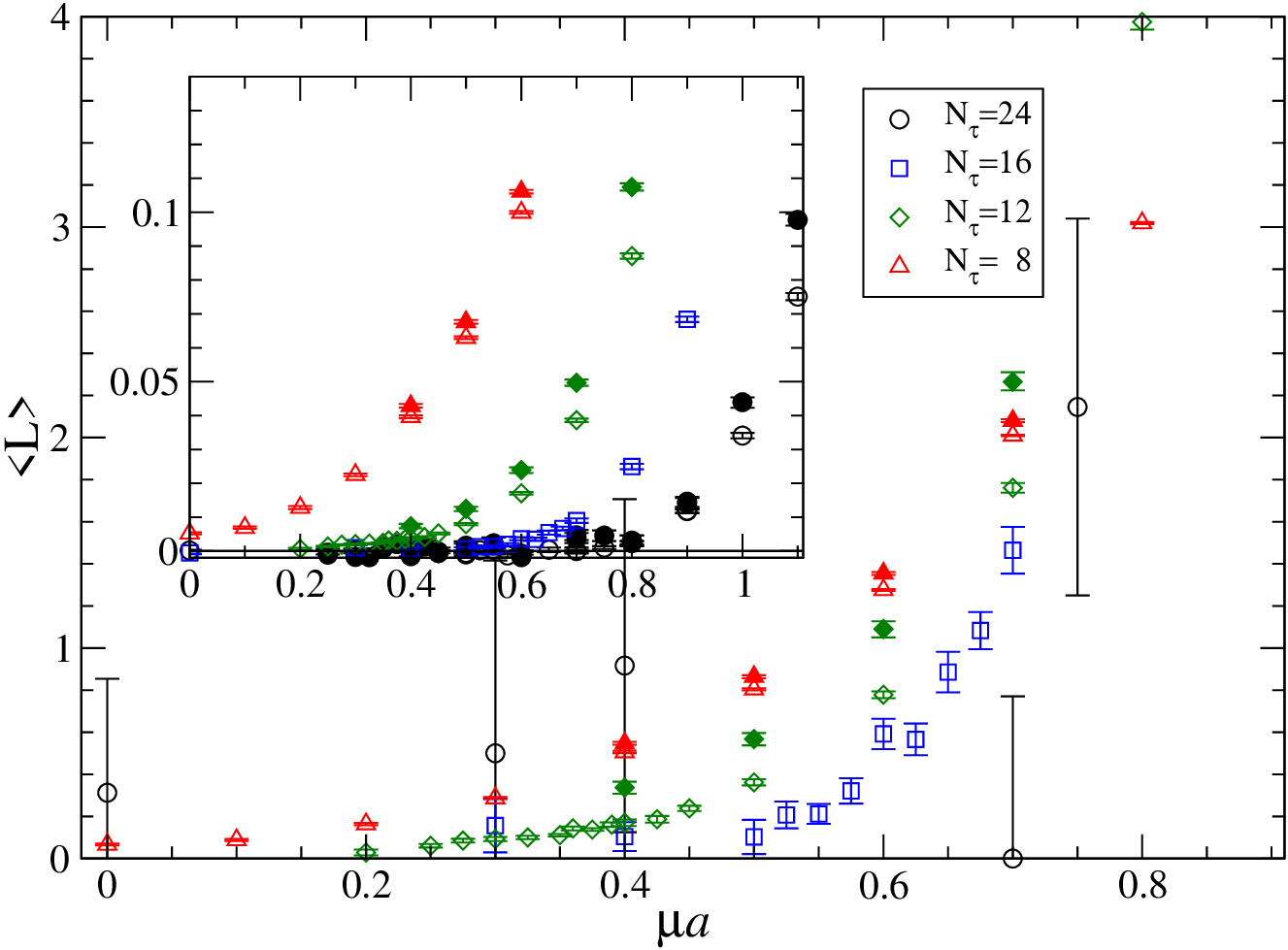}
\caption{\label{fig:polyakov} Polyakov line $L(\mu)$ for various $T$.}
\end{minipage}\hspace{2pc}%
\end{figure}
In order to characterise  the high density system we have also examined
two ``order parameters'', one exact, the other approximate. Fig.~\ref{fig:qq_j0}
shows the superfluid order parameter $\langle qq\rangle$ as a function of $\mu$,
extrapolated to $j=0$, for three different temperatures. For a degenerate system
superfluidity arises through Cooper pair condensation of diquark pairs at the
Fermi surface; hence $\langle qq\rangle$ should scale as the area of the Fermi
surface $k_F^2$. The plot shows $\langle qq\rangle/\mu^2$, and indeed it is
remarkably constant within the quarkyonic r\'egime, although clearly
$T$-sensitive. The temperature sensitivity becomes more marked once $\mu>\mu_d$,
and at the highest temperature studied the order parameter vanishes everywhere,
implying restoration of the ``normal'' phase.

We also studied the Polyakov loop $L$ as a function of both $\mu$ and $T$.
Since this involves comparison of data with different $N_\tau$, it
has been necessary to renormalise $L$ via multiplication by a factor
$Z_L^{N_\tau}$~\cite{Z_L}, determined at $\mu=0$ and normalised so that $L(T=1/4a)\equiv1$.
The results are shown in Fig.~\ref{fig:polyakov}; the inset shows the
unrenormalised data. First consider the data from the lowest temperature with
$N_\tau=24$. Although in the presence of fundamental matter $L$ is not an exact
order parameter for global Z$_{N_c}$ center symmetry, its behaviour strongly
suggests that the transition at $\mu_d$ is for all practical purposes identical 
with deconfinement, ie. the free energy for a static fundamental color
source becomes finite once $\mu>\mu_d$. The quark density at deconfinement is
16 -- 32 fm$^{-3}$ (the uncertainty arises from the difficulty in
handling lattice artifacts discussed in the previous section), some 30 -- 60
times that of nuclear matter.

\begin{figure}[h]
\begin{minipage}{18pc}
\includegraphics[width=18pc]{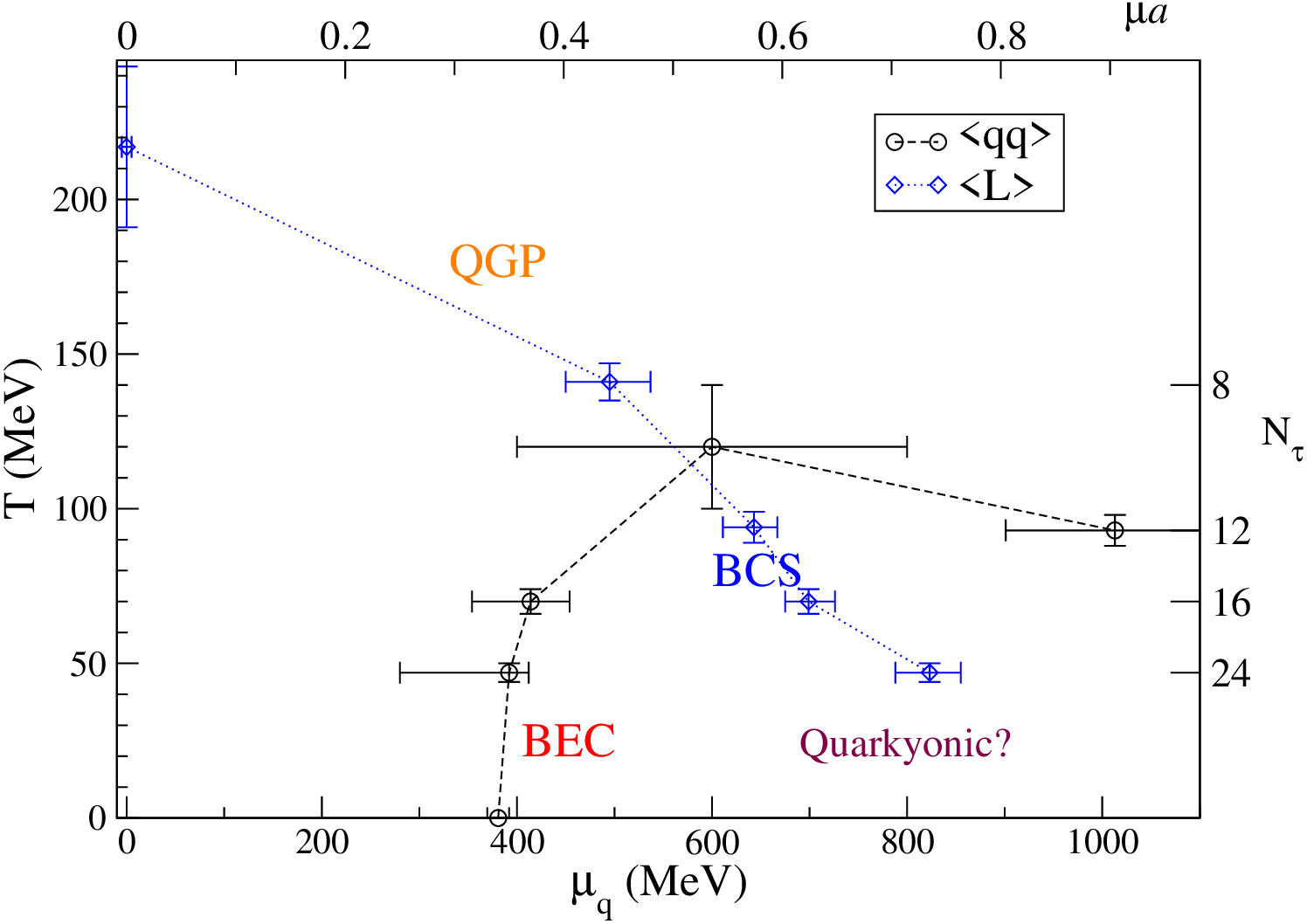}
\caption{\label{fig:phase} Tentative QC$_2$D phase diagram.}
\end{minipage}\hspace{2pc}%
\begin{minipage}{18pc}
\includegraphics[width=17pc]{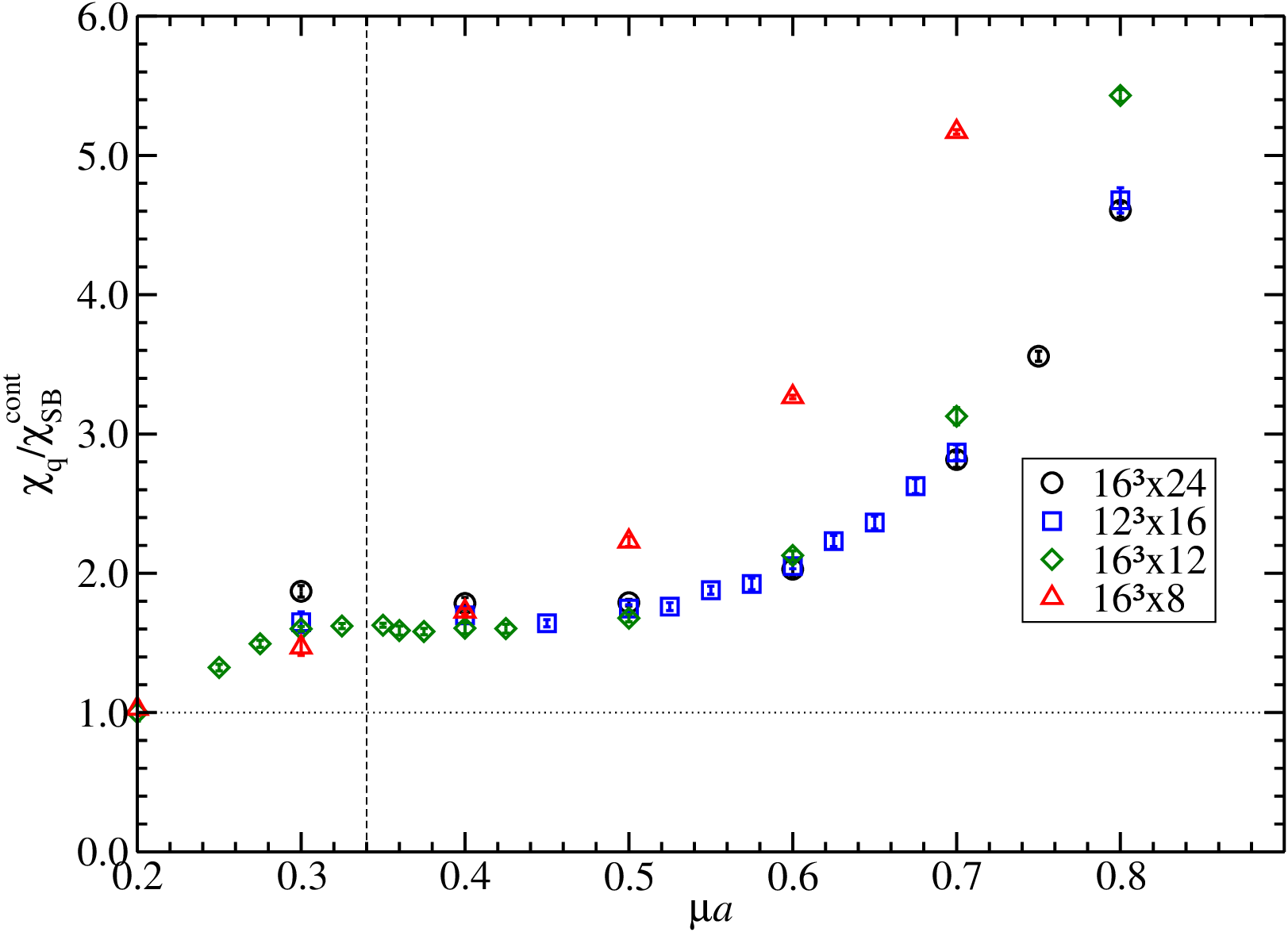}
\caption{\label{fig:chiq} $\chi_q/\chi_q^{SB}$ for $ja=0.04$, various $T$.}
\end{minipage} 
\end{figure}
It's then interesting that
$\mu_d$ defined via $L$ falls rapidly as $T$ rises. A pragmatic definition of
$\mu_d(T)$ is the value at which $L(\mu,T)\approx L(0,T_d)$.
The resulting tentative phase diagram is shown in
Fig.~\ref{fig:phase}. There are at least three distinct regions/phases: a normal
hadronic phase with $\langle qq\rangle=0$, $L\approx0$ at low $T$ and $\mu$; the
quarkyonic region with $\langle qq\rangle\propto\mu^2$ and $L\approx0$ at low
$T$ and intermediate $\mu$; and a deconfined, normal phase with $\langle
qq\rangle=0$, $L>0$ at large $T$ and/or large $\mu$. At this stage we cannot
exclude a deconfined superfluid phase at large $\mu$ and small/intermediate $T$.
The ``quarkyonic'' nomenclature is now justified; this phase has the
thermodynamic bulk scaling of a weakly-interacting degenerate quarks, but remains
confined. It thus has two of the features of dense baryonic matter originally proposed  on the basis
of large-$N_c$ arguments~\cite{quarkyonic}.

\section{Quark Number Susceptibility and Deconfinement}

We have also recently calculated 
the quark number susceptibility $\chi_q\equiv\partial
n_q/\partial\mu$~\cite{chiq,QC2D3}. Whilst
naively $\chi_q$ is expected to reflect local fluctuations of the
$n_q$ operator, it
turns out that at large $\mu$ the dominant term comes from a connected ``hairpin''
diagram. 
Fig.~\ref{fig:chiq} shows $\chi_q$ divided by the continuum
SB result. Once again, we note a large range over which 
the ratio is approximately constant; indeed it is compatible with one 
as $j\to0$, though more sensitive to the quark mass value used for
the free fermions than other bulk
observables~\cite{QC2D3}. 

The most striking feature of Fig.~\ref{fig:chiq},
however, is the absence of $T$-dependence at all but the highest temperature
studied (141 MeV), is in stark contrast to the behaviour of $L$ in
Fig.~\ref{fig:polyakov}. Despite our intuition from the thermal transition in
QCD~\cite{hotQCD}, $\chi_q$ cannot be regarded as a proxy for $L$ once $\mu/T\gg1$.
We conjecture that the change in behaviour of the bulk thermodynamic quantities 
$n_q$, $p$ and $\chi_q$ observed at $\mu_d$ is a transition from short-ranged
binary confining interactions to longer-ranged interactions among several quarks
within the medium, and that this corresponds with the transition from weak to
strong self-binding noted in Sec.~\ref{sec:eos}. The $T$-dependent behaviour of
$L$, in contrast, must be due to degrees of freedom close to the Fermi surface
which can be thermally excited; a
correlation between gapless excitations and $L>0$ has also been noted in
analytical and numerical studies on small cold volumes~\cite{attoworld}.
The same phenomenon should inform a theory of transport properties.

\section{Summary}
QC$_2$D offers an accessible theoretical laboratory for the study of dense
baryonic matter. For low $T$ at least three physical regions can be
identified: the vacuum for $\mu<\mu_o$; a confined quarkyonic superfluid for
$\mu_Q<\mu<\mu_d$; and a deconfined phase for $\mu>\mu_d$. The most recent
simulations have only made our findings in the quarkyonic r\'egime more robust.
Not discussed here are new results for renormalised energy density,
conformal anomaly, and chiral symmetry restoration~\cite{QC2D3}. 

\ack
This work used 
the DiRAC Facility jointly funded by 
STFC, the Large Facilities Capital Fund of BIS and 
Swansea University. We thank the DEISA Consortium 
(www.deisa.eu), funded through the EU FP7 project 
RI-222919, for support within the DEISA Extreme Computing
Initiative. 

\end{document}